\documentstyle[preprint,aps]{revtex}
\input{epsf}

\begin{document}

\tighten            

\title{Fragment Kinetic Energies and Modes of Fragment Formation}

\author
{
T.~Odeh,$^{(1)}$\cite{AAA}
R.~Bassini,$^{(2)}$
M.~Begemann-Blaich,$^{(1)}$
S.~Fritz,$^{(1)}$
S.J.~Gaff-Ejakov,$^{(3)}$
D.~Gourio,$^{(1)}$
C.~Gro\ss,$^{(1)}$
G.~Imm\'{e},$^{(4)}$
I.~Iori,$^{(2)}$
U.~Kleinevo\ss,$^{(1)}$\cite{AAA}
G.J.~Kunde,$^{(3)}$\cite{BBB}
W.D.~Kunze,$^{(1)}$
U.~Lynen,$^{(1)}$
V.~Maddalena,$^{(4)}$                   
M.~Mahi,$^{(1)}$
T.~M\"ohlenkamp,$^{(5)}$
A.~Moroni,$^{(2)}$
W.F.J.~M\"uller,$^{(1)}$
C.~Nociforo,$^{(4)}$                    
B.~Ocker,$^{(6)}$
F.~Petruzzelli,$^{(2)}$
J.~Pochodzalla,$^{(1)}$\cite{CCC}
G.~Raciti,$^{(4)}$
G.~Riccobene,$^{(4)}$                   
F.P.~Romano,$^{(4)}$                    
A.~Saija,$^{(4)}$                       
M.~Schnittker,$^{(1)}$
A.~Sch\"uttauf,$^{(6)}$\cite{DDD}
C.~Schwarz,$^{(1)}$
W.~Seidel,$^{(5)}$
V.~Serfling,$^{(1)}$
C.~Sfienti,$^{(4)}$                     
W.~Trautmann,$^{(1)}$
A.~Trzcinski,$^{(7)}$
G.~Verde,$^{(4)}$
A.~W\"orner,$^{(1)}$
Hongfei~Xi,$^{(1)}$
and B.~Zwieglinski$^{(7)}$
}

\address
{
$^{(1)}$Gesellschaft f\"ur Schwerionenforschung, D-64291 Darmstadt, 
Germany\\
$^{(2)}$Istituto di Scienze Fisiche, Universit\`{a} degli Studi
di Milano and I.N.F.N., I-20133 Milano, Italy\\
$^{(3)}$Department of Physics and
Astronomy and National Superconducting Cyclotron Laboratory,
Michigan State University, East Lansing, MI 48824, USA\\
$^{(4)}$Dipartimento di Fisica dell' Universit\`{a}
and I.N.F.N.,
I-95129 Catania, Italy\\
$^{(5)}$Forschungszentrum Rossendorf, D-01314 Dresden, Germany\\
$^{(6)}$Institut f\"ur Kernphysik,
Universit\"at Frankfurt, D-60486 Frankfurt, Germany\\
$^{(7)}$Soltan Institute for Nuclear Studies,
00-681 Warsaw, Hoza 69, Poland
}
\maketitle

\begin{center}

\end{center}

\begin{abstract}
Kinetic energies of light fragments ($A \le$ 10) from the
decay of target spectators in $^{197}$Au~+~$^{197}$Au collisions at 
1000 MeV per nucleon have been measured with high-resolution 
telescopes at backward angles. Except for protons and apart from the
observed evaporation components, the kinetic-energy spectra
exhibit slope temperatures of about 17 MeV, 
independent of the particle species, but not corresponding to the 
thermal or chemical degrees of freedom at breakup. It is suggested 
that these slope temperatures may reflect the intrinsic Fermi motion
and thus the bulk density of the spectator system at the instant of
becoming unstable.
\vspace{1.0cm}

\noindent
PACS numbers: 25.70.Pq, 21.65.+f, 25.70.Mn

\end{abstract}

\newpage

Statistical multifragmentation models have been found to be remarkably
successful in describing the partition space populated in nuclear 
reactions leading to multi-fragment emission \cite{gross90,bond95,hirsch99}. 
This is particularly true for spectator decays following heavy-ion 
collisions at relativistic bombarding energies. 
Fragment yields and fragment correlations have been 
reproduced with high accuracy 
\cite{botv92,barz93,bao93,konop93,zheng96,schuet,xi97,lauret98,avde98,radu99},
including their dispersions around the mean behavior \cite{botv95}.
The model temperatures are in good agreement with measured isotope 
temperatures, and the isotopic yields from which these temperatures are 
derived are well reproduced \cite{xi97,radu99}.

The mean kinetic energies of the produced particles and fragments,
on the other hand, are not universally accounted for if the same set 
of model parameters is used \cite{schuet,xi97}.
Measured kinetic energies or, equivalently, the slope temperatures 
obtained from Maxwell-Boltzmann fits to the kinetic energy spectra
have been found to considerably exceed the predicted values. This holds
even though the effects of Coulomb repulsion and sequential decay cause the 
calculated slope temperatures to be higher than the equilibrium temperatures 
in the model description. For the $^{197}$Au + $^{197}$Au reaction studied
in this paper, the measured isotope temperatures and the calculated 
equilibrium temperatures are 6 to 8 MeV for most of the impact 
parameter range; the calculated slope 
temperatures for light particles are about 10 MeV while the measured
slope temperatures exceed 15 MeV \cite{xi97}.

The qualitative difference between the chemical or thermal equilibrium 
temperatures and the kinetic temperatures has been observed for a variety 
of reactions
and numerous reasons for it have been presented
\cite{hirsch84,poch87,mori94,fuchs94,bauer95}.
Preequilibrium or pre-breakup emissions are likely explanations for
high-energy components in light-particle 
spectra \cite{botv92,xi97,morl96,wang99}. In some cases, the
excess kinetic energies have been ascribed to collective flow 
even though a characteristic proportionality to the fragment mass 
is not evident in the data \cite{lauret98,haug98,gula99}.
Other authors were caused by this situation
to more generally question the applicability of statistical descriptions
for spectator fragmentation \cite{goss97}. An explanation in terms of the 
Goldhaber model \cite{gold74} that is favored here has been suggested long
ago \cite{hirsch84,huef85} but, so far, has not been generally adopted.

In this Letter, we analyze kinetic-energy spectra
for fragment and light-charged-particle
emission from target spectators following collisions of 
$^{197}$Au + $^{197}$Au at 1000 MeV per nucleon. 
We will show that the deduced slope temperatures 
and the much lower chemical breakup temperatures
can be consistently understood 
if the intrinsic constituent motion of the decaying fermionic system
is taken into account. With this interpretation, the slope temperatures
reflect the bulk density of the system prior to its disintegration,
and thus provide information on the mechanism of fragment formation.

Beams of $^{197}$Au with incident energy 1000 MeV per nucleon
from the heavy-ion synchrotron SIS were used to bombard $^{197}$Au targets 
of 25-mg/cm$^2$ areal thickness.
As part of a larger experimental setup \cite{xi97,serf98,fritz99},
three high-resolution telescopes were placed at backward angles
for detecting the products of the target-spectator decay.
Each telescope consisted of 
three Si detectors with thickness 50, 300, and 1000 $\mu$m and 
of a 4-cm long CsI(Tl) scintillator with photodiode readout
and subtended a solid angle of 7.0 msr.

The products of the projectile decay were measured with the time-of-flight 
wall of the ALADIN spectrometer \cite{schuet} and $Z_{bound}$
was determined event-by-event. 
The sorting variable $Z_{bound}$ is defined
as the sum of the atomic numbers $Z_i$ of all projectile fragments
with $Z_i \geq$ 2. $Z_{bound}$ reflects the variation of the charge of 
the primary spectator system and is therefore correlated with the 
impact parameter of the reaction. Because of the
symmetry of the collision system, the mean values of $Z_{bound}$
for the target and the projectile spectators 
within the same event class have been assumed to be identical.
First results of these measurements have been presented in
Ref. \cite{xi97}, including some of the kinetic-energy data that 
will be discussed in the following.

Energy spectra of protons, measured at $\theta_{lab}$ = 150$^{\circ}$
and sorted into eight bins of $Z_{bound}$, are shown in Fig. 1 (left panel).
They are characterized by a hard component and an additional soft component 
that is most clearly identified at large $Z_{bound}$.
Fits using two Maxwellians yield temperature parameters 
of very different magnitude that both are nearly independent
of $Z_{bound}$ (Fig. 1, right panels).
The slope temperature $T$ of about 5 to 7 MeV of the soft 
component is of the same order as the measured breakup temperatures
\cite{xi97}. Its intensity, at the same time, is strongly correlated
with that of heavier fragmentation products \cite{odeh99}
which suggests a, perhaps partial, interpretation
as evaporation from highly excited residual nuclei. 
For the $^{197}$Au + $^{12}$C reaction at the same energy,
an equilibrium proton component with nearly identical properties was
reported by Hauger {\it et al.} \cite{haug98}.

Two components have also been identified in the spectra 
of $\alpha$ particles (Fig. 2) and in the neutron data measured with
LAND \cite{gross98}, with low-temperature components typical for
evaporation. 
The high-temperature components of protons and neutrons
seem to have their origin not only in the breakup stage but
also in the earlier cascading stages of 
the collision; their slope parameters are rather large and, as 
established for the neutron case \cite{gross98}, vary with the
bombarding energy.

The high-temperature component of $\alpha$ particles decreases from
$T \approx$ 17 MeV for small $Z_{bound}$ to $T \approx$ 13 MeV for
large $Z_{bound}$ (see below).
Its mean value of 15 MeV is close to the slope temperatures observed
for other species with $A \ge$ 2 which exhibit Maxwellian spectra
to a good approximation (Fig. 2). Their slope temperatures were
obtained from single-component fits with three-parameter functions 
that included a Coulomb potential $V_c$. The results for the interval
20 $\le Z_{bound} \le$ 60 are shown in Fig. 3.
Apart from the protons with $T \approx$ 26 MeV, 
all spectra exhibit temperatures that are narrowly dispersed around a
mean value of 17 MeV.

A mass-invariant temperature, 
at first sight, seems to provide direct evidence for equilibration 
of the kinetic degrees of freedom. 
On the other hand, Coulomb effects should contribute in proportion 
to the fragment charge, recoil effects may be important for heavier 
fragments, and the small 
but finite motion of the target spectator should introduce a collective 
velocity component. In order to estimate the magnitude of these effects, 
a model study was performed which started from 
initial configurations generated by randomly placing fragments and 
light particles in a spherical volume with random momenta corresponding to 
a given temperature.
Experimental charge distributions from Ref. \cite{schuet}
and an average density $\rho /\rho_0$ = 0.3 were used where 
$\rho_0$ is the normal nuclear density. 
N-body Coulomb trajectory calculations were then
performed and asymptotic slope temperatures were 
determined by fitting, following the same procedure as with the 
experimental data.
The calculations showed that the considered effects are relatively small, 
perhaps of the order of $\Delta T/\Delta A \approx$ 0.2 MeV, and that
they cancel each other to a good approximation at the 
backward angles chosen for the measurement. The measured slope
parameters of $T \approx$ 17 MeV (Fig. 3) are thus equal to 
the temperatures to be used in a thermal interpretation of the data.

The model study also showed that the Coulomb repulsion contributes 
considerably to the final kinetic energies. 
The resulting mean kinetic energies in the center-of-mass frame 
of the decaying system add up to about 35 to 40 MeV, quite consistent 
with values obtained in earlier experiments where kinetic energies 
were derived from the widths of momentum distributions of 
projectile fragments \cite{schuet}. In these experiments, the 
mean kinetic energies were found to be approximately constant over the 
range of fragment atomic number $Z~\le$~20.

It was suggested many years ago by Goldhaber that the product
momenta in fast fragmentation processes 
may have their origin in the nucleonic
Fermi motion within the colliding nuclei \cite{gold74}. He also pointed
out that the resulting behavior is indistinguishable from that
of a thermalized system with rather high temperature.
For the Fermi momentum $p_F \approx$ 265 MeV/c of heavier nuclei the
corresponding temperature is $T$ = 15 MeV. This is intriguingly
close to the measured slope temperatures $T$ = 17 MeV.

Goldhaber's idea, initially formulated for cold nuclei,
has been extended to the case of expanded
fermionic systems at finite temperature by Bauer \cite{bauer95}.
We have used the numerical solution reported there by assuming
that the temperature of the system is given by
the measured breakup temperatures $T_{\rm HeLi}$, derived from the
yields of $^{3,4}$He and $^{6,7}$Li isotopes \cite{xi97,albergo,poch95}.
For the breakup density two alternative values $\rho /\rho_0$~= 1.0 
and 0.3 were chosen which correspond to two significantly different 
scenarii for the process of fragment production.
The higher value is expected for a fast abrasion stage that will
induce fragmentation of the residual spectators
but not instantly affect the nucleon motion within them.
The lower value is in the range usually assumed for
multi-fragment breakups of expanded systems \cite{gross90,bond95}. 
Here the development of instabilities, 
as the system enters the spinodal region, will equally cause a rapid
fragmentation, so that the nucleonic Fermi motion will
contribute to the fragment kinetic energies. 
For a homogeneous system at a lower than normal density
the Fermi motion is reduced and the effect will be smaller.

The slope temperatures obtained in this way 
are represented by the lines shown in Fig.~4. The comparison
with the data is here restricted to the $A \le$ 4 isotopes for which 
the collected statistics are sufficient for studying 
the $Z_{bound}$ dependence. The recoil factor $(A_s-A)/(A_s-1)$ 
appearing in the Goldhaber formula \cite{bauer95,gold74}
can be safely ignored 
as the mass number $A_s$ of the spectator system is 100 to 150 on 
average and still about 50 in the bin of smallest $Z_{bound}$ \cite{poch95}.
Qualitatively, the predicted values are close to those observed.
In particular, the rise with decreasing $Z_{bound}$ follows as a 
consequence of the rising breakup temperatures $T_{\rm HeLi}$.
Apparently, with assumptions as made in the Goldhaber model, 
the two different types of temperatures are mutually consistent.
The role of the slope temperatures, consequently, is restricted
to describing the fragment motion while
$T_{\rm HeLi}$ is the more suitable observable
for representing the temperature of the nuclear environment at the breakup 
stage.

Recent calculations with transport models, which incorporate Fermi motion,
support this interpretation \cite{goss97,gait99}. 
The energies of spectator fragments, as measured in the present 
reaction, are well reproduced, and the coexistence of qualitatively 
different internal (or local) temperatures and fragment slope 
temperatures has been demonstrated in these
QMD and BUU studies.
On the experimental side, the observed 
similarity of fragment kinetic energy spectra, 
measured with different projectiles at various incident energies, 
is expected from the Goldhaber model
\cite{schuet,hirsch84,gula99,huef85,sang87,bert87,rensh96}. 
The fluctuations corresponding to the higher slope temperatures may also 
explain the observed large widths of the intrinsic kinetic-energy 
distributions in $^{197}$Au fragmentation at 600 MeV per nucleon that
are not easily reproduced with statistical multifragmentation 
models \cite{bege98}.

A more quantitative comparison with the data shown in Figs. 3 and 4 
will favor the prediction for $\rho /\rho_0$ = 1.0 corresponding to
a fast breakup over that for $\rho /\rho_0$ = 0.3 
that is reached after a homogeneous expansion.
This would not be unexpected in the present case of spectator fragmentation
following collisions of heavy ions and might indicate a limited 
resilience of nuclear matter to shape deformations \cite{colo97}.
The cited QMD and BUU calculations consistently suggest that the fragments 
are preformed at an early stage of the collision ($\le 50$ fm/c) before
the system has expanded to typical breakup densities.

To draw firm conclusions at this time would certainly be premature. 
The role of secondary decays and possibly other effects 
would have to be considered, and a moderate collective 
velocity component as a result of thermal expansion
cannot be completely excluded \cite{bege98,lefort99}, even though it is not 
explicitly indicated for the present reaction (Fig. 3 and \cite{schuet}).
The main purpose of this work is to show that, within the Goldhaber 
picture of a random superposition of the nucleon momenta, the larger
slope temperatures are fully consistent with a breakup at equilibrium 
temperatures of 6 to 8 MeV, as they are measured and also 
obtained with statistical 
multifragmentation models. Because of the sensitivity 
to the mode of fragment formation, via the density at which the Fermi
motion has been established, this
interpretation seems rather attractive and deserves further attention.

The authors would like to thank J. Aichelin, C. Fuchs, T. Gaitanos and H.H. 
Wolter for fruitful discussions. 
M.B., J.P., and C.S. acknowledge the financial support
of the Deutsche Forschungsgemeinschaft under the Contract No. Be1634/1-1,
Po256/2-1, and Schw510/2-1, respectively.
This work was supported by the European Community under
contract ERBFMGECT950083.

\begin{figure}[tb]
     \epsfysize=14.0cm
     \centerline{\epsffile{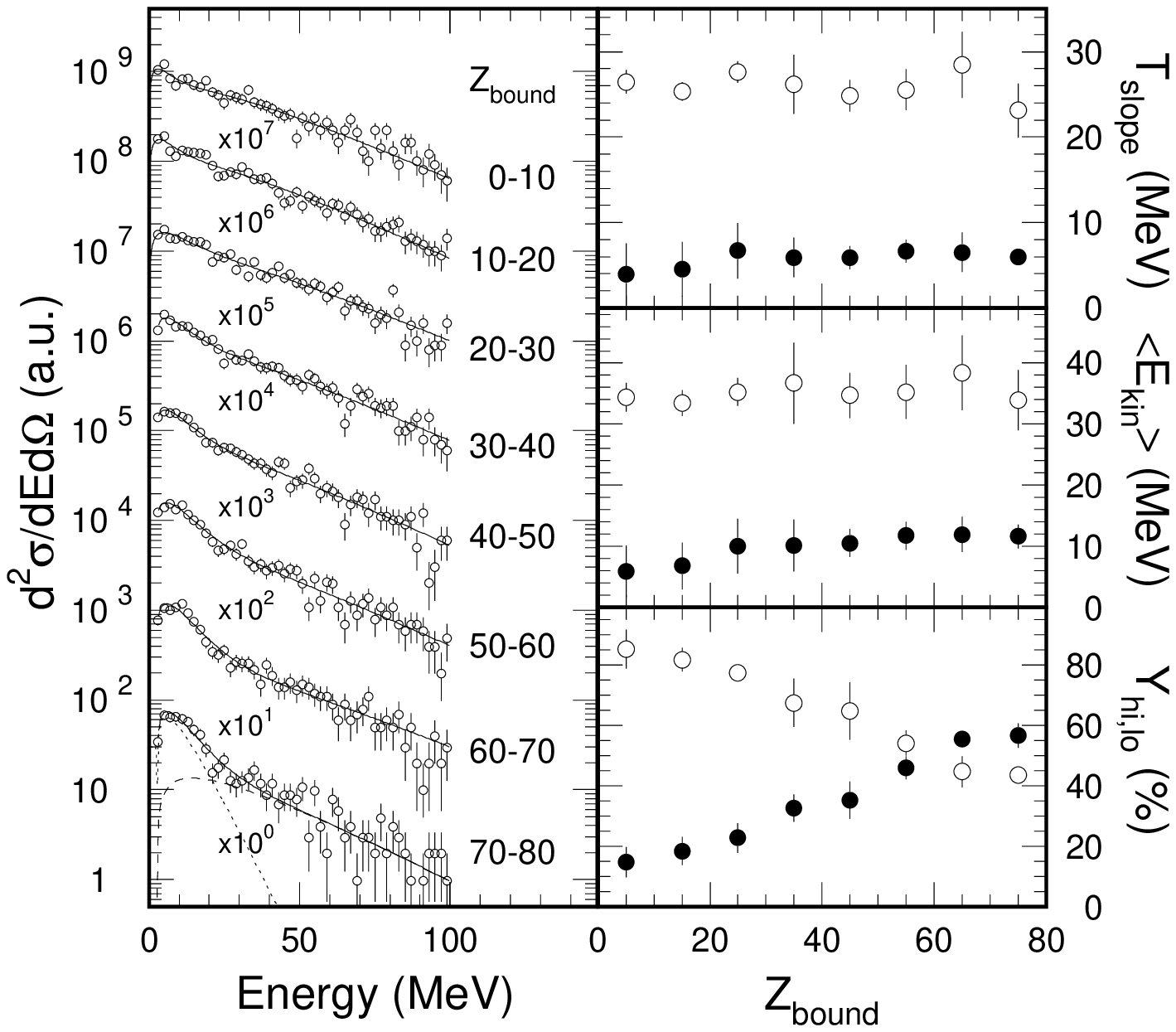}}
\vspace{1.0cm}

     \caption[]{
Proton energy spectra measured at $\theta_{lab}$ = 150$^{\circ}$
(left panel) and slope temperatures, mean kinetic energies, and relative 
yields of the two components with high (open circles) and low 
(closed circles) temperatures (right panels).
The full lines represent the fit results, the dashed and dotted lines
give the high and low temperature components 
for 70 $\le Z_{bound} \le$ 80 individually.
Note that the fitting was performed with the assumption that two 
components exist which, for $Z_{bound} \le$ 30, is not unambiguously 
supported by the data.
}
\label{FIG1}
\end{figure}

\begin{figure}[bbt]
     \epsfysize=14.0cm
     \centerline{\epsffile{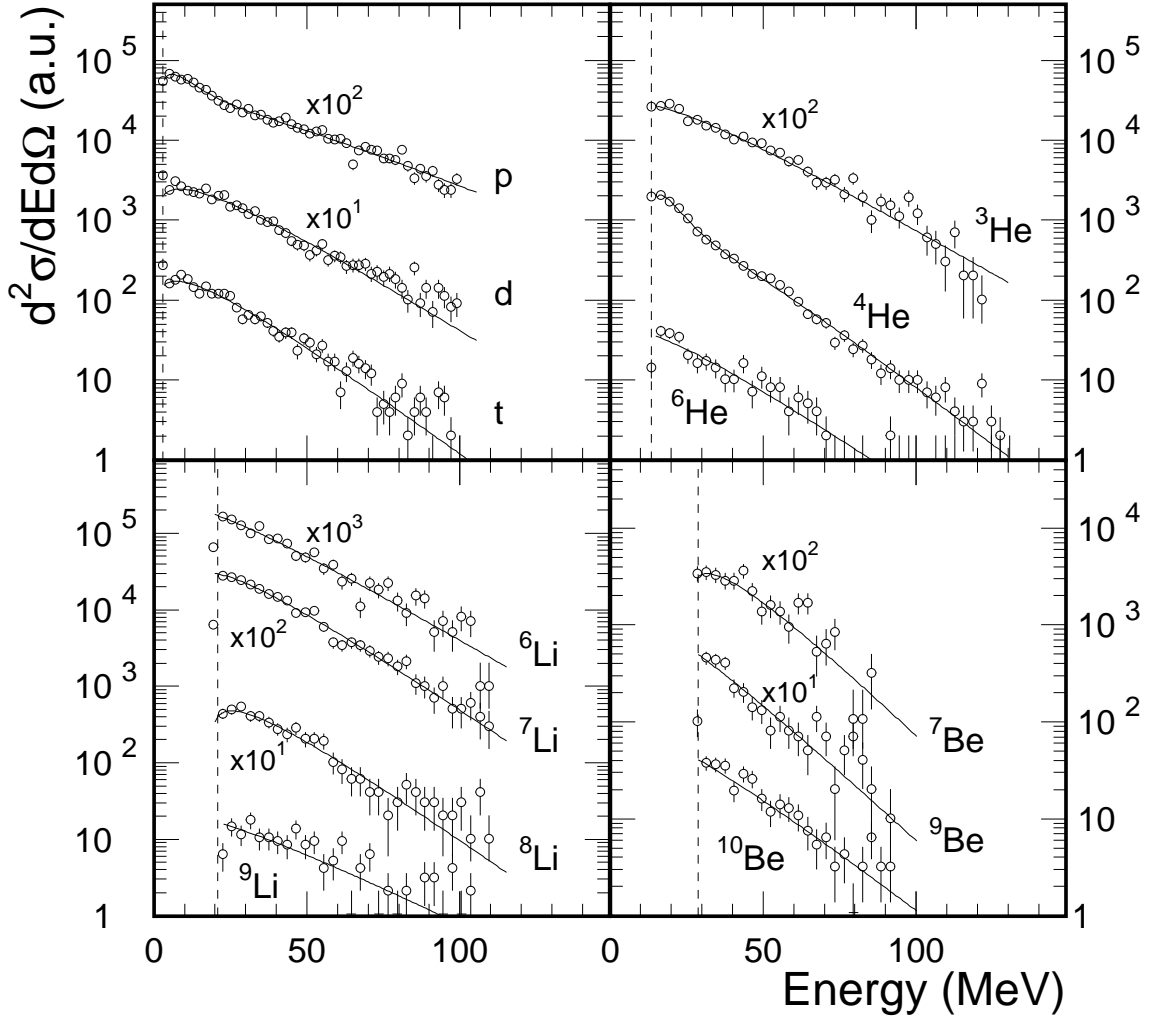}}
\vspace{1.0cm}

     \caption[]{
Energy spectra of light charged particles and fragments with $Z \le$ 4
integrated over 20 $\le Z_{bound} \le$ 60 ($\theta_{lab}$ = 150$^{\circ}$).
The full lines represent the fit results, the dashed lines indicate the
trigger threshold of the 300-$\mu m$ detector. The same normalization is
used for all spectra.
}
\label{FIG2}
\end{figure}

\begin{figure}[tb]
     \epsfysize=14.0cm
     \centerline{\epsffile{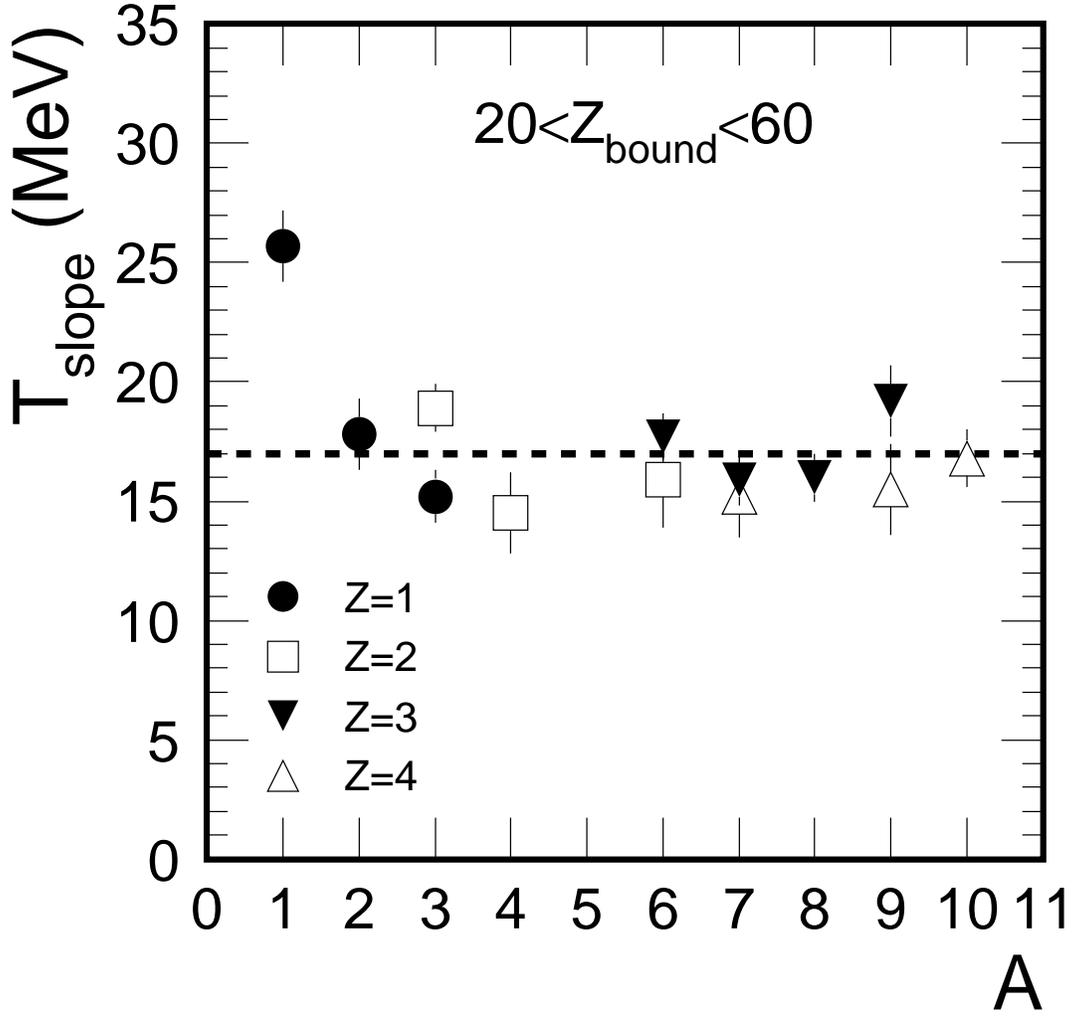}}
\vspace{1.0cm}

     \caption[]{
Slope temperatures for isotopically resolved charged particles and
fragments, measured at
$\Theta_{lab} = 150^{\circ}$ and integrated over 20 $\le Z_{bound} \le$ 60,
as a function of the mass number $A$.
For protons and $\alpha$ particles the value of the high-temperature 
component is shown.
}
\label{FIG3}
\end{figure}

\begin{figure}[tb]
     \epsfysize=14.0cm
     \centerline{\epsffile{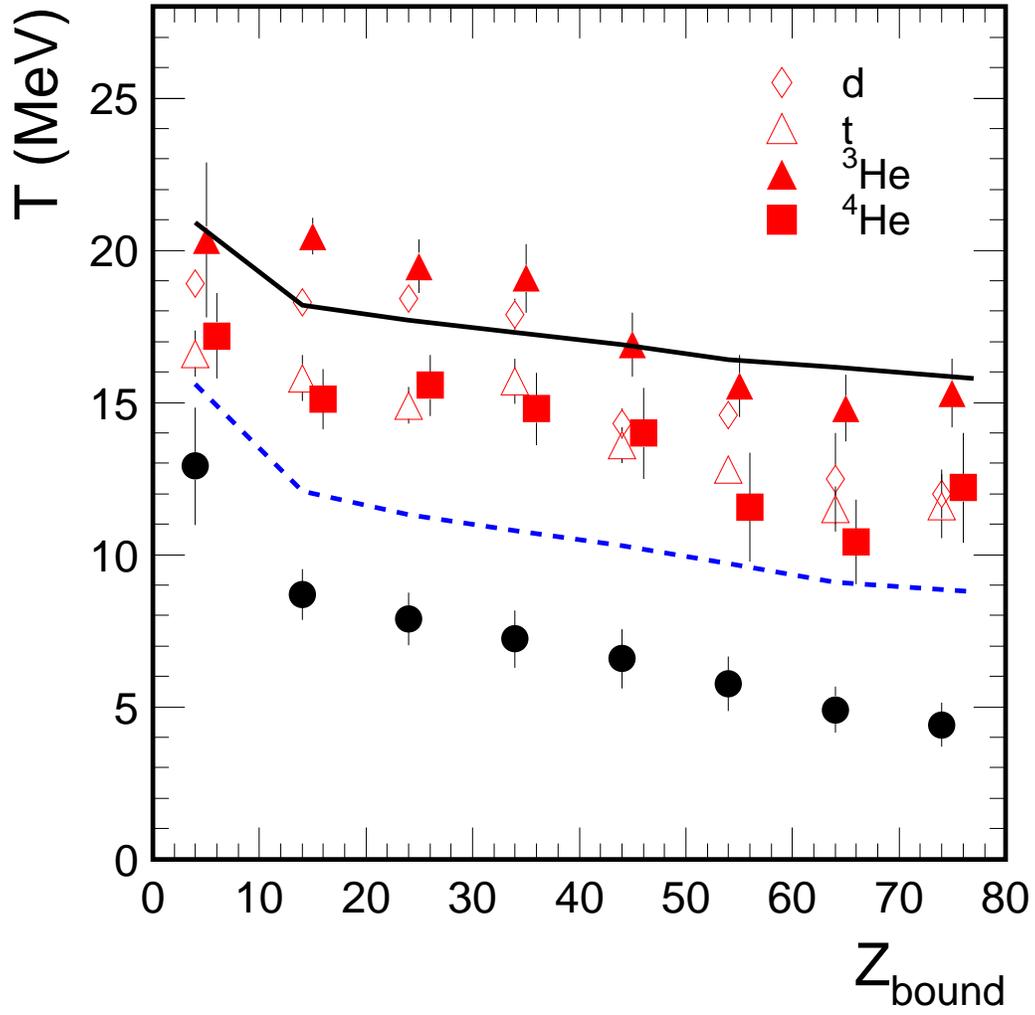}}
\vspace{1.0cm}

     \caption[]{
Slope temperatures for light charged particles of mass 2 $\le A \le$ 4
(squares, triangles and diamonds) and the 
isotope temperature $T_{\rm HeLi}$
(dots) as a function of $Z_{bound}$. The lines give the predictions
for the fast breakup of a Fermi gas
with finite temperature $T_{\rm HeLi}$ and with densities $\rho /\rho_0$ = 1.0
(full line) and 0.3 (dashed).
}
\label{FIG4}
\end{figure}

\end{document}